\documentstyle[11pt]{article}
\title{Stability of a Hot Smoluchowski Fluid}
\author{R. F. Streater,\\Dept. of Mathematics,\\King's
College,\\Strand, London WC2R 2LS.\\ray.streater@kcl.ac.uk\\http://
www.mth.kcl.ac.uk/$^\sim$streater/}
\date{26/10/2000}
\begin{document}
\maketitle
\begin{abstract}
We study coupled nonlinear parabolic equations for a fluid described by a
material density $\rho$ and a temperature $\Theta$, both functions of space
and time. In one dimension, we find some stationary solutions
corresponding to fixing the temperature on the boundary, with no-escape
boundary conditions for the material. For the special case, where the
temperature on the boundary is the same at both ends,
the linearized equations for small perturbations about
a stationary solution at uniform temperature and density are derived;
they are subject to boundary conditions, Dirichlet for $\Theta$ and
no-flow conditions for the material.

The spectrum of the generator $L$ of time evolution, regarded as an
operator on $L^2[0,1]$, is shown to be real, discrete and non-positive,
even though $L$ is not self-adjoint.
This result is necessary for the stability of the stationary state, but
might not be sufficient. The problem lies in the fact that $L$ is not
a sectorial operator, since its numerical range is ${\bf C}$.
\end{abstract}
\section{Introduction}
In \cite{RFS1,RFS2,RFS3,RFS4,RFS5} we derived systems of parabolic partial
differential
equations as the diffusion limit of discrete stochastic models
describing a fluid of identical particles moving in an external potential
$V(x)$. In the continuum limit the material is described by a density
$\rho(x,t)$ and the internal energy is described by a temperature field
$\Theta(x,t)$. The various models are based on different kinematic
assumptions, and this shows up in the details of the equations of evolution.
In \cite{RFS1,RFS5} there is a maximum possible value of the density,
called $\rho_{\rm max}$, corresponding to the state in which all the sites
of the discrete model are occupied. We showed that the first and second
laws of thermodynamics hold in these models,
and that in each case, the system has the Onsager
form \cite{RFS1,RFS6}, though it is not linear and not near equilibrium.
The difference between \cite{RFS5} and \cite{RFS1} is that in the latter,
the hopping probability in the underlying discrete model grows proportionally
with the energy of the state the particle occupies. This assumption leads to
a surprising phenomenon in the continuum limit: the model shows the Soret
and Dufour effects. This had been derived in 1912 by Enskog \cite{Enskog}
and in 1912 by Chapman \cite{Chapman} from the
Boltzmann equation, but only for gas mixtures.

These and similar models have been extended and studied by
Biler, Dolbeault, Esteban, Hebisch, Karch, Krzywicki, Markowich and Nadzieja
in a series of papers \cite{BD,BDEK,BDM,BHN,BKN,BN1,BN2,BN3,KN,NR}. 
In these papers, $V$ is either an external potential
or is the mutual gravitational or electrical potential between the particles,
and so obeys a Helmholtz equation with $\rho(x,t)$ as its source.
This leads to an integro-differential system, for which the authors
have been able to prove existence and uniqueness of solutions in many
cases, and to study the asymptotic return to equilibrium.

In this paper we concentrate on the model \cite{RFS1}
with the Soret and Dufour effects.
The heat capacity per particle is unity, and so
the energy density is naturally defined to be
\begin{equation}
{\cal E}(x,t):=\rho(x,t)[\Theta(x,t)+V(x)].
\end{equation}
The dynamics conserves the total number of particles and the total energy,
and this leads to the local conservation laws:

\begin{eqnarray}
\frac{\partial\rho}{\partial t}&+&\mbox{div}\,j_c=0\label{material1}\\
\frac{\partial{\cal E}}{\partial t}&+&\mbox{div}\,j_e=0.\label{energy}
\end{eqnarray}
The material current $j_c$ and the energy current $j_e$ of the model are
shown to be
\begin{eqnarray}
j_c&=&-\lambda\left(\Theta\nabla\rho+\rho(1-\rho/\rho_{\rm max})
\nabla(\Theta+V)\right),\\
j_e&=&2\left(\Theta j_c-\lambda\rho(1-\rho/\rho_{\rm max})\Theta\nabla\Theta
\right)+Vj_c.
\end{eqnarray}

For rare gases, the density is much less than its maximum, and
the system of equations simplifies. The currents for
material and energy are then
\begin{eqnarray}
j_c&=&-\lambda\left(\Theta\nabla\rho+\rho\nabla(\Theta+V)\right)
\label{matcurrent}\\
j_e&=&2\left(\Theta j_c-\lambda\rho\Theta\nabla\Theta\right) +Vj_c.
\label{encurrent}
\end{eqnarray}
The derivation given in \cite{RFS1} starts with a Markov chain, and
takes the continuum limit. The equations are therefore local in time, and
if $V$ is time-dependent, the same calculation shows that the right-hand
side of eq.~(\ref{energy}) should be $\rho\partial V/\partial t.$
The model falls into the general scheme advocated by Wojnar \cite{Wojnar}.

In \cite{RFS1} we show that the entropy is an
increasing function of time (except at a stationary configuration);
thus the model obeys both the first and the second law of thermodynamics
if the boundary conditions allow no
material or heat to enter or leave the system. The entropy is shown to be
\begin{eqnarray}
S&=&-\int\rho(x)\,\log\rho(x)\,dx-\int(\rho_{\rm max}-\rho(x))\log(1-\rho/
\rho_{\rm max})\,dx+\nonumber\\
&+&\int\rho(x)\log\Theta(x)\,dx.
\end{eqnarray}
For small $\rho(x)/\rho_{\rm max}$, so that we can ignore $(\rho/
\rho_{\rm max})^2$, this simplifies to
\begin{equation}
S=-\int\rho(x)\,\log\rho(x)\,dx+\int\rho(x)\,\log\Theta(x)\,dx
\end{equation}
apart from the term $\int\rho(x)\,dx$, which is constant in time
if no material enters or leaves the system.

Usually, the energy-balance equation (\ref{energy}) is ignored,
and the temperature put equal to $\Theta_0$ for all space and time
(the iso-thermal situation), and we reduce to the Smoluchowski equation
for the Brownian particle in a potential. We avoid this
simplification, which loses the
Soret and Dufour effects that come from the cross-terms $\nabla\Theta$ in
eq.~(\ref{matcurrent}) and $2\Theta j_c$ in $j_e$. The first is the
Joule-Thompson effect (in the local form called the Soret effect), which
ensures that a gradient in the temperature causes a flow of material, even
when the density is constant. The term $2\Theta j_c$ is the Dufour effect,
which means that there is a flow of heat even when the temperature is
constant, if there is a non-zero particle current. We might expect the flow
of heat due to convection to be
$\Theta j_c$ since the heat capacity is unity, and the factor
$2$ was a surprise. In \cite{RFS1} it was shown
that the other half of the contribution $2\Theta j_c$ is needed
as the Onsager dual of the Soret term, $\nabla\Theta$, which occurs
in $j_c$.

In this paper, we study these equations in one dimension in a bounded
interval, which without loss of generality may be taken to be $[0,1]$.
We shall continue to write $\nabla$ for the gradient, so for the rest
of the paper, $\nabla=\partial/\partial x$.

Consider the the boundary conditions $\Theta=\Theta_0$ for all time at $x=0$
and $x=1$, and $j_c=0$ for all time at $x=0$ and  $x=1$.
These have the direct physical meaning that the system is in a heat-bath
at temperature $\Theta_0$, and no material escapes through the boundary,
which is impermeable to the particles of the fluid, but conducts heat
perfectly. Such mixed Dirichlet-Neumann conditions for systems do not appear
to have been considered before in that it leads to a new type of
generator $L$ of the semigroup of time-evolution, for which form techniques
do not work. This is because the operator $L$ is not sectorial.

It is obvious that there is a stationary
solution satisfying the following conditions: $\Theta(x,t)=\Theta_0$,
and $j_c(x,t) =0$ for all space and time. It is then easy to solve for
$\rho$ in the case
of a constant gravitational field $(\S2)$. We also briefly discuss the case
when $\Theta$ is not constant.

In \S3 we study the stability of small perturbations about
the stationary state, by the usual method.
We limit the
consideration to the case $V=0$, and so leave for the future the
(B\'{e}nard) instabilities expected for a liquid in a gravitational field
heated from below and cooled from above.

We give a proof that the eigenvalues are real and non-positive,
by exhibiting a $2\times 2$ similarity transform (not unitary) that
transforms the operator $L$ to a diagonal positive matrix, times
the known negative self-adjoint operator, the Laplacian with
self-adjoint boundary conditions.
This shows that $L$ is closed, and leads to the general study of operators of
this type, which has been pursued by L. Boulton. The new difficulty is that
the numerical range of the generator is the whole complex plane.

\section{Some No-flow Solutions}
Let $j_c$ and $j_e$ be given by eq.~(\ref{matcurrent}) and
(\ref{encurrent}).
Let us impose the no-flow conditions $\mbox{div}\,j_c=0$, so in one dimension,
$j_c$ is a constant. This must be zero, since we impose the boundary
condition of no material flow at the boundary. Let us impose
Dirichlet conditions $\Theta(0)=\theta_0$ at $x=0$ and $\Theta(1)=\theta_1$
at $x=1$. The condition for
stationarity of the energy-density then leads to $\mbox{div}(\rho\Theta\nabla
\Theta)=0$; these two remarks can be expressed as
\begin{eqnarray}
(\Theta\rho)^\prime+\rho V^\prime&=&0\\
(\Theta\Theta^\prime\rho)^\prime&=&0.
\end{eqnarray}
The second can be solved exactly for $\rho$ in terms of $\Theta$.
We assume that $\Theta(x)$ and $\Theta(x)^\prime$
have no zeros in $[0,1]$. This gives
\begin{equation}
\rho=k/(\Theta\Theta^\prime),
\label{material2}
\end{equation}
where $k$ is the constant of integration, assumed not to be zero.
Eliminating  $\rho$ from the first equation, we get
\begin{equation}
\Theta\Theta^{\prime\prime}-\Theta^\prime V^\prime=0
\label{first}
\end{equation}

This can be solved exactly for $\Theta$ if V(x) is linear, and
$\theta_0=\theta_1$. Thus suppose that
\begin{equation}
V(x)=gx
\end{equation}
as in a constant gravitational field. Then we get the obvious solution
$\Theta(x) =\theta_0$={\rm const.}, and this leads to the solution for the
density:
\begin{equation}
\rho(x)=C\exp\{-gx/\Theta_0\}.
\end{equation}
This is just that given by the Boltzmann factor, exhibiting the usual
exponential fall-off of the density with height. This result shows
that the equations are on the right track. 

We can find many exact solutions by choosing $\Theta$, satisfying the
boundary conditions, instead of choosing $V$. We then just need to solve
the first-order equation (\ref{first}) for $V(x)$.
Not all choices of $\Theta(x)$ lead to nonsingular equations. For example,
if we try to satisfy boundary conditions in which $\theta_0=\theta_1$,
then a non-negative differentiable function $\Theta(x)$ must have a
zero for $\Theta^\prime$ somewhere in the interval $[0,1]$. Then both
$\rho$ and the force $-\nabla V$ must be infinite somewhere in the interval.
It follows that the only continuous stationary solution with $\theta_0=
\theta_1$, and $\Theta^\prime$ finite, is constant in $x$. It is easy
find stationary solutions satisfying other boundary conditions;
for example choose $\Theta(x)= (x+1)^2$; then we have a stationary solution
\begin{eqnarray}
\rho(x)&=&k(x+1)^{-3}\\
\Theta(x)&=&(x+1)^2\\
V(x)&=&x+x^2/2.
\end{eqnarray}
This satisfies the equations together with the conditions that $j_c(x)=0$
for all $x$, and the boundary conditions $\Theta(0)=1,\hspace{.in}
\Theta(1)=4$. The energy-flow, however is not zero, but is $-4\lambda$.
This flows in at $x=1$ and out at $x=0$, and is constant throughout
the region. We have a driven stationary system far from equilibrium.
We can follow the production of entropy explicitly; heat enters the
system at a high temperature, $\Theta=4$ and leaves at a low temperature,
$\Theta=1$. Heat leaves at the
same rate as it enters. So the system receives entropy at the rate
$Q/\Theta(1)=4\lambda/4=\lambda$, and ejects it at the rate $Q/\Theta(0)=
4\lambda$, a net production rate of $3\lambda$. Where is the location
of this entropy production?
If we use the Onsager formalism, we can write the entropy production
as \cite{RFS1}, equation (4.11)
\begin{equation}
\Theta\dot{S}=\int j_c.X^c/\Theta+\int j_e.X^e/\Theta.
\label{ent}
\end{equation}
Here, $X^c$ and $X^e$ are the thermodynamic forces, given by \cite{RFS1}
(4.13) and (4.14). We only need
\begin{equation}
X^e/\Theta=\nabla(1/\Theta)
\end{equation}
since $j_c=0$.
The entropy produced by the system is therefore
\begin{equation}
\dot{S}=\int_0^1 j_e.X^e/\Theta \,dx=-4\lambda[(x+1)^{-2}]_0^1=3\lambda.
\label{onsager}
\end{equation}
At a stationary distribution, there is no change in the
microscopic probability distribution with time, so it seems that the
von Neumann entropy (from which the entropy in \cite{RFS1} is derived)
must be independent of time. This is true, and is derived in
eq.~(4.10) in \cite{RFS1}.
In getting the Onsager form eq.~(\ref{ent}), we discarded the boundary
term, which if kept would appear as the {\em loss} of entropy by the system
at the rate $3\lambda$, due to receiving hot heat and emitting cooler heat.
The total change in the entropy is therefore zero, as befits a
time-independent probability distribution.
The environment, however, has an increase in entropy at the rate
of $3\lambda$.

It is easy to construct many more examples.

\section{Analysis of stability in the field-free case}
We limit this section to the study of the case, in one dimension, on
the interval $[0,1]$, where
$V=0$, and $\Theta(0)=\Theta(1)$. We choose to
write the dynamics in terms of the
fields ${\cal E}:=\rho\Theta$ and $\Theta$; these have simple boundary
conditions, Neumann and Dirichlet at the points $x=0$ and $x=1$,
respectively. Note that at the boundary, $\nabla {\cal E}=0$, because the
{\em material} current, which is $j_c=-\lambda\nabla {\cal E}$, vanishes
on the boundary. The energy current
is $2\Theta j_c-2\lambda\rho\Theta\nabla\Theta=-2\lambda\nabla(\Theta{\cal
E})$. We can recover $\rho={\cal E}/\Theta$ from these fields.

The equations of continuity, (\ref{material1}) and (\ref{energy})
then become
\begin{eqnarray}
\frac{\dot{\cal E}}{\Theta}-\frac{{\cal E}\dot{\Theta}}{\Theta^2}&=&
\lambda\nabla^2{\cal E}\\
\dot{\cal E}&=&2\lambda\nabla^2(\Theta{\cal E}).
\end{eqnarray}
These can be written
\begin{eqnarray}
\dot{\Theta}&=&\frac{2\lambda\Theta}{\cal E}(\Theta{\cal E})^{\prime\prime}
-\lambda\frac{\Theta^2}{\cal E}{\cal E}^{\prime\prime}\\
\dot{\cal E}&=&2\lambda(\Theta{\cal E})^{\prime\prime}.
\end{eqnarray}
These clearly have the stationary solution $\Theta(x)=\Theta(0)=\Theta_0$,
${\cal E}(x)={\cal E}(0)={\cal E}_0$, for all $x\in[0,1]$. To test
the stability of this solution, we put
\begin{equation}
\Theta(x,t)=\Theta_0 +\theta(x,t);\hspace{.5in}{\cal E}(x,t)
={\cal E}_0+e(x,t)
\end{equation}
where $\theta(x,t)$ and $e(x,t)$ are small functions, so that we ignore
powers higher than the first, of these and their derivatives.
Up to the first order, the equations become
\begin{eqnarray*}
\dot{e}&=&2\lambda\left[(\Theta_0+\theta)({\cal E}_0+e)\right]
^{\prime\prime}\\
&=&2\lambda\left[{\cal E}_0\theta^{\prime\prime}+\Theta_0e^
{\prime\prime}\right].
\end{eqnarray*}
Also
\begin{eqnarray*}
\dot{\theta}&=&\frac{2\lambda(\Theta_0+\theta)}{{\cal E}_0+e}\left[
(\Theta_0+\theta)({\cal E}_0+e)\right]^{\prime\prime}\\
& &-\frac{\lambda(\Theta_0+\theta)^2}{{\cal E}_0+e}e^{\prime\prime}\\
&=&\frac{2\lambda\Theta_0}{{\cal E}_0}\left[{\cal E}_0\theta^
{\prime\prime}+(\Theta_0/2)e^{\prime\prime}\right].
\end{eqnarray*}
The linearized version of the equations can therefore be summarised as
\begin{equation}
\left(\begin{array}{c}
\dot{e}\\
\dot{\theta}\end{array}\right)=2\lambda\Theta_0\left[\begin{array}{cc}
1&\gamma\\
1/(2\gamma)&1
\end{array}\right]\left(\begin{array}{c}
e^{\prime\prime}\\
\theta^{\prime\prime}
\end{array}\right)
:=L\left(\begin{array}{c}
e^{\prime\prime}\\
\theta^{\prime\prime}
\end{array}\right)
\label{eigen}
\end{equation}
where $\gamma={\cal E}_0/\Theta_0$.
The boundary conditions are $e^\prime(0)=e^\prime(1)=0=\theta(0)=
\theta(1)$. A necessary condition for the stability of
the solution $({\cal E}_0,\Theta_0)$ is that
the non-zero part of the spectrum of the operator
\begin{equation}
M\otimes\frac{d^2}{dx^2}=\left[\begin{array}{cc}
1&\gamma\\
1/(2\gamma)&1
\end{array}
\right]\otimes\frac{d^2}{dx^2}
\end{equation}
entering eq.~(\ref{eigen})
lies in the open left-half plane. We shall prove this, and show that the
spectrum is discrete and negative, of finite multiplicity. Zero
is in the spectrum, since $e(x)=$ non-zero constant, $\theta(x)=0$
is the corresponding eigenfunction. This perturbation corresponds to 
moving from one stationary state to another, with a different total number
of particles.

$M$ is not symmetric, but is diagonalisable by a similarity,
by Sylvester's criterion.
In fact, there is a one-parameter family of diagonalising matrices; for,
if $SMS^{-1}=\widehat{M}$ is diagonal, then the one-parameter family of
diagonal matrices $B(\beta)={\rm diag}[\beta,1/\beta]$
commute with $\widehat{M}$, and so leave it diagonal: $B\widehat
{M}B^{-1}=\widehat{M}$. In general, such a transformation leaves us with
non-self-adjoint boundary conditions. These can be expressed as
a singular perturbation of the Laplacian, in the manner of \cite{albeverio}.
This monograph, however, only deals with the self-adjoint case.
This extra one-parameter family of diagonalising
similarities can be used to ensure that the boundary conditions define
a symmetric Dirichlet form, by a making judicious choice of $\beta$.
Note that it is enough to diagonalise
$(M-I)$; this leads to the special case, $\alpha=1/2$, of the
problem of finding all matrices $S$ such that
\[
S=\left[\begin{array}{cc}
a&b\\
c&d
\end{array}\right]\]
and
\begin{equation}
S\left[\begin{array}{cc}
0&\gamma\\
\alpha/\gamma&0
\end{array}\right]S^{-1}={\rm diag}[\lambda_1,\lambda_2]
\end{equation}
Direct calculation immediately gives
the conditions for the off-diagonal terms in $\widehat{M}$ to vanish:
\begin{eqnarray*}
b^2\alpha&=&a^2\gamma^2\\
\alpha d^2&=&c^2\gamma^2.
\end{eqnarray*}
To avoid the vanishing of det\,$S$ we have to choose opposite signs
in the square-roots of these equations, for example
\begin{eqnarray*}
b&=&a\gamma\alpha^{-1/2}\\
d&=&-c\gamma\alpha^{-1/2}.
\end{eqnarray*}
Then $\widehat{M}={\rm diag}[1+\alpha^{1/2},1-\alpha^{1/2}]$.
We notice that this is independent of the value of $\gamma$.
If we can show that the operator $\widehat{M}\otimes d^2/dx^2$
is a Dirichlet form,
it is negative since $\alpha=1/2<1$. For this to be the case, the boundary
conditions must be self-adjoint, so that the boundary terms that arise
in the integration by parts vanish for functions in the domain of
the operator. Let
\[
\left(\begin{array}{c}
\hat{e}\\
\hat{\theta}
\end{array}\right)=S\left(\begin{array}{c}
e\\
\theta
\end{array}\right),\]
where
\[
S=\left(\begin{array}{cc}
a&a\gamma\alpha^{-1/2}\\
c&-c\gamma\alpha^{-1/2}
\end{array}\right).
\]
Thus,
\begin{eqnarray*}
\hat{e}&=&ae+a\gamma\alpha^{-1/2}\theta\\
\hat{\theta}&=&ce-c\gamma\alpha^{-1/2}\theta.
\end{eqnarray*}
Then
\begin{eqnarray*}
(\hat{e},\hat{\theta})\hat{M}\left(\begin{array}{c}
\hat{e}^{\prime\prime}\\
\hat{\theta}^{\prime\prime}
\end{array}\right)&=&(1+\alpha^{1/2})\langle\hat{e},\hat{e}^{\prime\prime}
\rangle+(1-\alpha^{1/2})\langle\hat{\theta},\hat{\theta}^
{\prime\prime}\rangle\\
&=&-(1+\alpha^{1/2})\langle\hat{e}^\prime,\hat{e}^\prime\rangle-
(1-\alpha^{1/2})\langle\hat{\theta}^\prime,\hat{\theta}^\prime\rangle\\
& &+(1+\alpha^{1/2})\left[\hat{e}\hat{e}^\prime\right]_0^1+(1-\alpha^{1/2})
\left[\hat{\theta}\hat{\theta}^\prime\right]_0^1.
\end{eqnarray*}
The boundary terms which must vanish are thus
\begin{eqnarray*}
& &(1+\alpha^{1/2})\left[(ae+a\gamma\alpha^{-1/2}\theta)(ae^\prime+a\gamma
\alpha^{-1/2}\theta^\prime)\right]_0^1+\\
& &+(1-\alpha^{1/2})\left[(ce-c\gamma\alpha^{-1/2})(ce^\prime-c
\gamma\alpha^{-1/2}\theta^\prime)\right]_0^1\\
&=&\left\{(1+\alpha^{1/2})\gamma\alpha^{-1/2}a^2-(1-\alpha^{1/2})
\gamma\alpha^{-1/2}c^2\right\}\left[e\theta^\prime\right]_0^1.
\end{eqnarray*}
For this to vanish for all $e,\theta$ in the domain, we need to be able
to choose $(1+\alpha^{1/2})a^2=(1-\alpha^{1/2})c^2$ without making
$S$ singular; this is possible. We conclude that our operator $L$
is similar to
a negative self-adjoint operator with compact resolvent, and so has
non-positive discrete spectrum. The domain of the operator is
\begin{equation}
\{(e,\theta)
\in W^{2,2}:e^\prime(0)=e^\prime(1)=0\mbox{ and }\theta(0)=\theta(1)=0\}.
\end{equation}
In order to solve the eigenvalue problem, we try
\begin{eqnarray*}
\hat{e}(x,t)&=&A\exp\{-\omega t\}\cos(k_1x)+B\exp\{-\omega t\}\sin(k_1x)\\
\hat{\theta}(x,t)&=&C\exp\{-\omega t\}\cos(k_2 x)+D\exp\{-\omega t\}
\sin(k_2x).
\end{eqnarray*}
We have proved above that $\omega$, an eigenvalue of $-L$, is nonnegative;
$k_1$ and $k_2$ must be chosen to satisfy
\begin{equation}
\frac{\omega}{2\lambda\Theta(0)}=(1+\alpha^{1/2})k_1^2=(1-\alpha^{1/2})k_2^2.
\end{equation}
To fit the boundary conditions, we can write $(\hat{e},\hat{\theta})$ in
terms of $(e,\theta)$ and then use the given conditions. However,
since a similarity does not alter the eigenvalue (only the eigenfunctions)
we can avoid the complication of choosing $S$ so that the boundary terms
vanish, and instead use any convenient matrix that diagonalises $M$.
We choose
\begin{equation}
S=\left(\begin{array}{cc}
1&\alpha^{-1/2}\\
-1&\alpha^{-1/2}
\end{array}\right).
\end{equation}
This gives us the boundary conditions
\begin{eqnarray}
\hat{e}^\prime-\hat{\theta}^\prime&=&0,\hspace{.4in} \mbox{at }x=0,\;x=1\\
\hat{e}+\hat{\theta}&=&0,\hspace{.4in}\mbox{at }x=0,\;x=1.
\end{eqnarray}
By imposing this condition, we get a linear homogeneous system of equations
for $A,B,C,D$, and the vanishing of the determinant gives us
a holomorphic trancendental equation for $\omega$. Note that the eigenvalues
do not depend on $\gamma$, the ratio of ${\cal E}_0$ to $\Theta_0$.
We have computed the eigenvalues $\omega$ of $-L$, and indeed, they
are all non-negative.

In \cite{Kato} it is proved that the semigroup $U(t)=\exp\{Lt\}$
defined by a sectorial
operator $L$ with range in the left half-plane is a contraction semigroup
in $L^2$. Here, however, the conditions for the theorem fail, and the best
we can do in $L^2$ is to get the inequality for $u$ orthogonal to
the one-dimensional subspace of stationary solutions:
\begin{equation}
\|\exp\{Lt\}u\|_2\leq \exp\{-\lambda_1t\}\|S\|\;\|S^{-1}\|\;\|u\|_2
\end{equation}
where $\lambda_1$ is the smallest non-zero eigenvalue of $-L$. Since
$\|S\|\;\|S^{-1}\|>1$ in our model, the system is not obviously
stable for small $t$. This problem does not arise for an isolated system,
in which there is no flow of heat or material at the boundary; this
gives us Neumann conditions on the boundary for both $\nabla{\cal E}$
and $\nabla\Theta$, which is a self-adjoint
condition. Then the numerical range of $M\otimes d^2/dx^2$ is
determined by its symmetric part, which is non-positive.

\vspace{.2in}
\noindent{\bf Acknowledgements}

\vspace{.1in}
\noindent The author would like to thank E. B. Davies and L. Boulton for
very useful discussions, and T. Nadzieja for reading the manuscript.

\end{document}